\documentclass[showpacs]{revtex4}
\usepackage{makeidx}
\usepackage{amsfonts}
\usepackage{amsmath}
\usepackage{latexsym}
\usepackage{epsfig}

\begin{document}

\title{Anomalous Transport and Nonlinear Reactions in Spiny Dendrites}
\author{Sergei Fedotov$^1$, Hamed Al-Shamsi$^1$, Alexey Ivanov$^2$ and
Andrey Zubarev$^{2}$}
\date{\today }

\begin{abstract}
We present a \textit{mesoscopic }description of the anomalous transport and
reactions of particles in spiny dendrites. As a starting point we use
two-state Markovian model with the transition probabilities depending on
residence time variable. The main assumption is that the longer a particle
survives inside spine, the smaller becomes the transition probability from
spine to dendrite. We extend a linear model presented in [PRL, \textbf{101},
218102 (2008)] and derive the nonlinear Master equations for the average
densities of particles inside spines and parent dendrite by eliminating
residence time variable. We show that the flux of particles between spines
and parent dendrite is not local in time and space. In particular the
average flux of particles from a population of spines through spines necks
into parent dendrite depends on chemical reactions in spines. This memory
effect means that one can not separate the exchange flux of particles and
the chemical reactions inside spines. This phenomenon does not exist in the
Markovian case. The flux of particles from dendrite to spines is found to
depend on the transport process inside dendrite. We show that if the
particles inside a dendrite have constant velocity, the mean particle's
position $\left\langle x(t)\right\rangle $ increases as $t^{\mu }$ with $\mu
<1$ (anomalous convection). We derive a fractional convection-diffusion
equation for the total density of particles.
\end{abstract}

\affiliation{$^1$School of Mathematics, The University of Manchester, Manchester M60 1QD,
UK}
\affiliation{$^2$ Department of Mathematical Physics, Ural State University, Russia }
\maketitle

\section{Introduction}

Dendritic spines play a very important role in regulating neuronal activity
of the cerebellar cortex because the majority of excitatory synapses are
located on spines \cite{Sabatini0,Segal}. The later are tiny bulbous
protrusions from dendrites consisting of head ($\sim 1$ $\mu $m)\ and thin
neck ($\sim 0.1$ $\mu $m). One of the main functions of spines is to help
transmit electrical signals. The problem of propagation of action potential
in the spiny dendrites has attracted enormous attention in past years.
Several cable models have been suggested in the literature to study the
spine-dendrite interaction on the \textit{macroscopic} level. Baer and
Rinzel \cite{BR} developed a phenomenological cable theory for spiny
dendrites and found that the propagation rate of local excitation strongly
depends on a spine-stem resistance. The dynamic structure of spines (the
changes in the shape and size) has been studied in \cite{WuBaer}. Modified
FitzHugh-Nagumo model \cite{coom} was the subject of research by Coombes and
Bressloff. Population of spines can be treated as a continuous function of
spatial position along the dendrite \cite{EB,Ti}. It should be noted that
these cable models are phenomenological and not derived from the
electro-diffusion equations for ions in spiny dendrites.

Recent experiments together with numerical simulations by Santamaria \textit{%
et al} \cite{san} showed that the transport of inert particles in spiny
dendrites is anomalous. It was found that the mean-square displacement $%
\left\langle x^{2}(t)\right\rangle $ is proportional to $t^{\mu }$ with $\mu
<1$. It turns out the dendritic spines act as traps of particles. The narrow
spine neck significantly decreases the effective diffusion of particles into
and out of dendritic spines. This results in slow anomalous diffusion along
the shaft of dendrite \cite{mk}. Based on these experiments, Henry \textit{%
et al} \cite{Henry} suggested the fractional Nernst-Planck equations for
electro-diffusion of ions in spiny dendrites and developed a cable model
involving time-fractional derivatives. In particular, the average flow of
ions between the shaft of dendrite and the spines is not local in time. Note
that the coupling between spines and dendrites has been studied either
phenomenologically or on \textit{microscopic} level of a single spine \cite%
{Svoboda,Schuss,Ber}. The influence of a dendritic spine on the spread of
calcium in the parent dendrite was studied in \cite{KS}. The passive
diffusion in a tube with dead ends has been considered in \cite{Ber2}.
Recently \textit{mesoscopic} non-Markovian model for spines-dendrite
interaction has been developed \cite{FM1,MFH}. The aim of this paper is to
give an alternative Markovian model involving the residence time variable.
This paper is an essential extension of \cite{FM1} with new results and
examples.

We are concerned with a \textit{mesoscopic }description of the anomalous
transport and reactions of particles in spiny dendrites. The \textit{%
mesoscopic} approach involves a detailed description of the behavior of
particles on the \textit{microscopic} level. At the same time within \textit{%
mesoscopic} approach we can introduce the mean densities of particles and
neglect the random fluctuations around the mean behavior \cite{MFH}. The
system of linear Master equations for the mean density of particles inside a
dendrite, $n_{1}(x,t),$ and the density of particles inside a population of
spines, $n_{2}(x,t)$ has been derived in \cite{FM1}. The starting point was
non-Markovian continuous time random walk (CTRW) model involving the
integral balance equations for average densities. The aim of this paper is
to extend this analysis for non-linear case and derive a mesoscopic system
of equations from Markovian model. One way to deal with non-Markovian
process is to introduce supplementary variables that make the process
Markovian \cite{Cox,Kampen}. Here we introduce the residence time variable $%
\tau $ which is the time interval between arrival of particle in dendrite or
spine and the time of leaving them. The particle has a zero residence time
when it just arrives inside the spines from dendrite or inside dendrite from
spines. This idea has been used in \cite{Cox,VR,YH,Nep}.

\section{Two-state reaction-transport model}

We start with the \textit{microscopic} two-state model of a random
particle's movement in a spiny dendrite. When the particle is inside a
dendrite, it performs a random walk for a random time $T_{1}$ (dendrite's
residence time) before hitting the neck of spine. When the particle hits the
neck, it is trapped inside the spine for a random time $T_{2}$ (spine's
residence time). After spending the random time $T_{2}$ inside spine, the
particle is released to the parent dendrite through spine neck. The particle
starts to perform a random walk inside the dendrite until it hits the
spine's neck again. Clearly the essential feature of this process can be
described by two-state random process. The particle can be in one of the two
states: inside dendrite or inside spines. We assume that the residence times
or waiting times $T_{1}$ and $T_{2}$ are non-negative random variables with
probability density functions (pdf's) $\psi _{1}\left( \tau \right) $ and $%
\psi _{2}\left( \tau \right) $ respectively. Note that the introduction of
the residence time pdf's can be viewed as the probabilistic method to model
the variability in the shape of dendritic spines and their density on the
parent dendrite. The corresponding survival functions $\Psi _{i}(\tau )$ are
defined as
\begin{equation}
\Psi _{i}(\tau )=\int_{\tau }^{\infty }\psi _{i}(s)ds\qquad i=1,2.
\label{sur}
\end{equation}%
To describe the switching process we introduce two transition probabilities
(hazard functions) \cite{Cox}:
\begin{equation}
\gamma _{i}(\tau )=\frac{\psi _{i}(\tau )}{\Psi _{i}(\tau )},\qquad i=1,2
\label{hazard}
\end{equation}%
Thus $\gamma _{1}(\tau )h$ represents the conditional probability of
transition from dendrite to spine in the small interval $(\tau ,\tau +h)$
given that there is no transition up to time $\tau .$ The product $\gamma
_{2}(\tau )h$ has a similar meaning for the transition from spine to
dendrite. We also assume that inside spines the irreversible chemical
reaction $C\overset{r_{2}^{-}}{\rightarrow }C_{b}$ with the rate $r_{2}^{-}$
takes place. It describes the removal of the particles by immobile buffers
and pumps \cite{Svoboda,Schuss}.

If we assume that the transition probabilities $\gamma _{1}(\tau )$ and $%
\gamma _{2}(\tau )$ are constants, then we have a classical two-state
Markovian model. Master equations for the mean density of particles inside a
dendrite, $n_{1}(x,t),$ and the density of particles inside a population of
spines, $n_{2}(x,t),$ are
\begin{equation}
\frac{\partial n_{1}}{\partial t}=L_{x}n_{1}-\gamma _{1}n_{1}+\gamma
_{2}n_{2},  \label{mean1}
\end{equation}%
\begin{equation}
\frac{\partial n_{2}}{\partial t}=-r_{2}^{-}\left( n_{2}\right) n_{2}-\gamma
_{2}n_{2}+\gamma _{1}n_{1},  \label{mean2}
\end{equation}%
where the reaction rate $r_{2}^{-}\left( n_{2}\right) $ depends on the local
density of particles $n_{2}$ \cite{MFH}. Here $L_{x}$ is the transport
operator acting on $x$-coordinate along the dendrite.

To define the transport operator $L_{x},$ one can use the continuous-time
random walk (CTRW) model in which the particles follow the path of the
compound Poisson process. The probability of a jump during time interval of
length $h$ is $\lambda (x)h+o(h)$ and jumps density function is $w(z)$ \cite%
{MFH}. Then
\begin{equation}
L_{x}n_{1}=--\lambda (x)n_{1}+\int_{\mathbb{R}}\lambda (z)n_{1}(z,t)w(x-z)dz.
\label{L2}
\end{equation}%
If we adopt the Nernst--Planck equation for the particles flow then
\begin{equation}
L_{x}n_{1}=-\frac{\partial (v(x,t)n_{1})}{\partial x}+D\frac{\partial
^{2}n_{1}}{\partial x^{2}},  \label{L1}
\end{equation}%
where $D$ is the diffusivity of the particles. The advection velocity $%
v(x,t)=-\mu \partial \phi /\partial x$, where $\mu $ $\ $is the mobility,
and the electrostatic potential $\phi $. Note that $v(x,t)$ depends on the
total concentration of particles (ions) through the Poisson equation.

Our aim now is to extend the Markovian model (\ref{mean1}) and (\ref{mean2})
to the two-state semi-Markov model \cite{MFH}. Applications of the
semi-Markov processes to chemical kinetics can be found in \cite{WQ}. Note
that stochastic two-state models with nonexponential waiting time
distributions occur in many areas of natural sciences. We mention the
stochastic resonance theory \cite{GH3}, two-state model for anomalous
diffusion \cite{Shushin}, two-state gating process for ion channels \cite{GH}%
, propagation of tumor cells \cite{FI07}, superdiffusion theory and random
walk with memory \cite{FMT}.

\section{ Markovian model involving a residence time variable}

The purpose of this section is to formulate the Markovian model for the
transport and reactions of particles inside spiny dendrites. If the
transition probabilities $\gamma _{1}(\tau )$ and $\gamma _{2}(\tau )$ are
not constants, it is convenient to introduce the densities of particles
depending on $\tau $ \cite{VR}. Let $\xi _{1}(x,\tau ,t)$ be the density of
particles at point $x$ at time $t$ whose residence time inside the dendrite
lies in the interval $\left( \tau ,\tau +d\tau \right) $. The corresponding
density of particles inside a population of spines is $\xi _{2}(x,\tau ,t)$.
Integration of $\xi _{i}(x,\tau ,t)$ over residence time variable $\tau $
gives the mean densities $n_{i}(t,x)$ at point $x$ at time $t$:
\begin{equation}
n_{i}(x,t)=\int_{0}^{\infty }\xi _{i}(x,\tau ,t)d\tau ,\qquad i=1,2.
\label{den0}
\end{equation}%
This model is similar to well-known age-structured models in which the
population density of individuals depends explicitly on the age $\tau $ \cite%
{M}. Of course, we should make a clear distinction between the residence
time $\tau $ since the last jump and the residence time of a particle from $%
t=0.$ The latter is not considered here.

The crucial question here whether the particles "remember" how long they
have been inside spines or dendrite. The experiments \cite{san} suggest the
existence of the memory effects which lead to the anomalous transport of
particles along the dendrite. To model these effects, we assume that the
transition probabilities\ $\gamma _{i}(\tau )$ depend on the residence time
variable $\tau $ \cite{Cox}. Thus the probability of a transition from a
dendrite to spines during a small time interval of length $h$ is $\gamma
_{1}(\tau )h+o(h)$, and the backward transition has the probability $\gamma
_{2}(\tau )h+o(h)$.

Since the average movement of particles inside the dendrite is governed by
the operator $L_{x}$, the balance equations for $\xi _{1}(x,\tau ,t)$ and $%
\xi _{2}(x,\tau ,t)$ are

\bigskip \textit{For a dendrite}
\begin{equation}
\frac{\partial \xi _{1}}{\partial t}+\frac{\partial \xi _{1}}{\partial \tau }%
=L_{x}\xi _{1}-\gamma _{1}\left( \tau \right) \xi _{1}.  \label{dif1}
\end{equation}%
\textit{For spines}
\begin{equation}
\frac{\partial \xi _{2}}{\partial t}+\frac{\partial \xi _{2}}{\partial \tau }%
=-\gamma _{2}\left( \tau \right) \xi _{2}-r_{2}^{-}\left( n_{2}\right) \xi
_{2}.  \label{dif2}
\end{equation}%
The derivation of (\ref{dif1}) and (\ref{dif2}) together with the transport
operator (\ref{L2}) is given in Appendix A. Note that this Markovian model
can be easily generalized to include various nonlinear terms.

Initial conditions are
\begin{equation}
\xi _{i}(x,\tau ,0)=n_{i}^{0}(x)f_{i}(\tau |x)\qquad i=1,2  \label{in}
\end{equation}%
where $n_{i}^{0}(x)$ is the initial densities of particles and $f_{i}(\tau
|x)$ is the conditional waiting time distribution for the particles inside
dendrite ($i=1$) and spines ($i=2$) at time $t=0$ with the property $%
\int_{0}^{\infty }f_{i}(\tau |x)d\tau =1$ \cite{VR,YH}.

Boundary conditions at $\tau =0:$

\textit{For a dendrite}%
\begin{equation}
\xi _{1}(x,0,t)=\int_{0}^{\infty }\gamma _{2}(\tau )\xi _{2}(x,\tau ,t)d\tau
.  \label{b1}
\end{equation}%
\textit{For spines}%
\begin{equation}
\xi _{2}(x,0,t)=\int_{0}^{\infty }\gamma _{1}(\tau )\xi _{1}(x,\tau ,t)d\tau
\label{b2}
\end{equation}%
\textit{\ } These boundary conditions have the following meaning. Particles
inside dendrite with the residence time $\tau =0$ at the point $x$ are
created with the rate $\gamma _{2}(\tau )$ (see (\ref{b1})). The density of
particles just arriving inside dendrite, $\xi _{1}(x,0,t),$ can be found by
integration of the product $\gamma _{2}(\tau )\xi _{2}(x,\tau ,t)$ over all
residence times. The density of particles just arriving inside spines $\xi
_{2}(x,0,t)$ can be found in a similar way (see (\ref{b2})).

Now we assume that the convective velocity $v(x,t)=v=const$ and the jump
rate $\lambda (x)=const.$ Using the method of characteristics and the
Fourier transform, we obtain the solutions to (\ref{dif1}) and (\ref{dif2})
(see Appendix B):

\textit{For a dendrite}
\begin{equation}
\xi _{1}(x,\tau ,t)=e^{-\int_{0}^{\tau }\gamma _{1}(s)ds}\int_{\mathbb{R}%
}\xi _{1}(z,0,t-\tau )p(x-z,\tau ))dz\qquad \tau <t,  \label{sol1}
\end{equation}%
\begin{equation}
\xi _{1}(x,\tau ,t)=e^{-\int_{\tau -t}^{\tau }\gamma _{1}(s)ds}\int_{\mathbb{%
R}}\xi _{1}(z,\tau -t,0)p(x-z,t))dz\qquad \tau >t.  \label{sol2}
\end{equation}

\textit{For spines}
\begin{equation}
\xi _{2}(x,\tau ,t)=\xi _{2}(x,0,t-\tau )e^{-\int_{0}^{\tau }\gamma
_{2}(s)ds-\int_{t-\tau }^{\tau }r_{2}^{-}\left( n_{2}(x,s)\right) ds}\qquad
\tau <t,  \label{sol3}
\end{equation}%
\begin{equation}
\xi _{2}(x,\tau ,t)=\xi _{2}(x,\tau -t,0)e^{-\int_{\tau -t}^{\tau }\gamma
_{2}(s)ds-\int_{0}^{t}r_{2}^{-}\left( n_{2}(x,s)\right) ds}\qquad \tau >t.
\label{sol4}
\end{equation}%
Here $p(x,t)$ is the Green function for the transport equation
\begin{equation}
\frac{\partial p}{\partial t}=L_{x}p  \label{kf}
\end{equation}%
with the initial condition%
\begin{equation}
p(x,0)=\delta (x).  \label{kf0}
\end{equation}%
In particular, when the advection velocity $v=const$ in (\ref{L1}) we have
\begin{equation}
p(x,t)=\frac{1}{\sqrt{4\pi Dt}}\exp \left[ -\frac{(x-vt)^{2}}{4Dt}\right] .
\end{equation}%
Note that (\ref{dif1}) or (\ref{dif2}) has different solutions for $\tau <t$
and $\tau >t.$ When $\tau <t,$ the solution (\ref{sol1}) describes the
evolution of particles that arrived inside dendrite from spines after $t=0.$
The formula (\ref{sol2}) gives the evolution of density of those particles
that were present inside dendrite at $t=0.$

The formulas (\ref{sol1}) and (\ref{sol2}) can be generalized for the case
when the advection velocity $v$ and jump rate $\lambda $ depend on space and
time. We can write
\begin{equation}
\xi _{1}(x,\tau ,t)=e^{-\int_{0}^{\tau }\gamma _{1}(s)ds}\int_{\mathbb{R}%
}\xi _{1}(z,0,t-\tau )p(x,\tau |z))dz\qquad \tau <t,
\end{equation}%
\begin{equation}
\xi _{1}(x,\tau ,t)=e^{-\int_{\tau -t}^{\tau }\gamma _{1}(s)ds}\int_{\mathbb{%
R}}\xi _{1}(z,\tau -t,0)p(x,t|z))dz\qquad \tau >t,
\end{equation}%
where the Green function $p(x,t|z)$ can be interpreted as the probability
density function for a particle which starts at point $z$ in a dendrite and
arrives at point $x$ at time $t$ without trapping in spines up to time $t$.
The pdf $p(x,t|z)$ obeys the equation $\partial p/\partial t=L_{x}p$ with
the initial condition $p(x,0|z)=\delta (x-z)$. Of course for nonhomogeneous
case such as (\ref{L2}) or (\ref{L1}) the explicit expression for $p(x,t|z)$
is not available.

The density $\xi _{i}(x,\tau ,t)$ can be also interpreted as the probability
density function of finding the particle inside the dendrite ($i=1)$ or
spines ($i=2$) at the point $x$ at time $t$ such that the residence time
lies in the interval $(\tau ,\tau +d\tau ).$ Since $\dot{\Psi}_{i}(\tau
)=\psi _{i}(\tau ),$ it follows from (\ref{hazard}) that the survival
function $\Psi _{i}(\tau )$ is
\begin{equation}
\Psi _{i}(\tau )=e^{-\int_{0}^{\tau }\gamma _{i}(s)ds}\qquad i=1,2.
\label{su2}
\end{equation}%
One can see that this exponential factor appears in the solutions (\ref{sol1}%
)-(\ref{sol4}). That is why (\ref{sol1})-(\ref{sol4}) have a very simple
probabilistic meaning of the law of total probability. Note that the
residence time's density $\psi _{i}(\tau )$ can be written in terms of the
transition rate $\gamma _{i}(\tau )$ as follows \cite{Cox}
\begin{equation}
\psi _{i}(\tau )=\gamma _{i}(\tau )e^{-\int_{0}^{\tau }\gamma
_{i}(s)ds}\qquad i=1,2.  \label{su3}
\end{equation}%
It is natural to assume that the longer a particle survives inside spine,
the smaller becomes the transition probability from spine to dendrite. In
this case the transition rate $\gamma _{2}(\tau )$ is a monotonically
decreasing function of residence time $\tau .$ For example, if
\begin{equation}
\gamma _{2}(\tau )=\frac{\mu }{\beta +\tau }  \label{res}
\end{equation}%
then the survival function
\begin{equation*}
\Psi _{2}(\tau )=\left( \frac{\beta }{\beta +\tau }\right) ^{\mu }.
\end{equation*}%
Thus the assumption of the dependence of transition rate $\gamma _{2}(\tau )$
on the residence time like (\ref{res}) leads to a power-law probability
density function of residence time inside spines
\begin{equation}
\psi _{2}(\tau )=\frac{d\Psi _{2}}{d\tau }=\frac{\mu }{\beta }\left( \frac{%
\beta }{\beta +\tau }\right) ^{\mu +1}.  \label{res1}
\end{equation}%
The experimental evidence to support the hypothesis of a power-law
distribution like (\ref{res1}) with $\mu <1$ is the subdiffusive transport
of particles in a spiny dendrite \cite{san,FM1}. In what follows we will use
this distribution to derive the fractional equations for the densities of
particles.

\section{Non-Markovian two-state model}

The aim of this section is to set up a non-Markovian model for the transport
and reactions of particles in spiny dendrites by eliminating the residence
time variable $\tau $. Let us denote the densities of particles just
arriving in a dendrite and spines at point $x$ at time $t$ by $j_{1}(x,t)$
and $j_{2}(x,t)$ respectively. Obviously
\begin{equation}
j_{i}(x,t)=\xi _{i}(x,0,t)\qquad i=1,2.  \label{den}
\end{equation}%
Note that in the paper \cite{FM1} $j_{1}(x,t)$ denotes the number of
particles arriving at point $x$ inside dendrite at time $t$ through a single
spine stem (not a population of spines considered here) and $j_{2}(x,t)$ is
the number of particles arriving at point $x$ in a single spine at time $t.$

To derive the balance equations for $j_{1}(x,t)$ and $j_{2}(x,t)$ we
substitute (\ref{sol1})-(\ref{sol4}) into the boundary conditions (\ref{b1}%
),(\ref{b2}) and use (\ref{in}),\ (\ref{su3}). It gives the following
equations for\ $j_{1}(x,t)$ and $j_{2}(x,t)$
\begin{eqnarray}
j_{1}(x,t) &=&\int_{0}^{t}\psi _{2}(\tau )j_{2}(x,t-\tau )e^{-\int_{t-\tau
}^{t}r_{2}^{-}\left( n_{2}(x,s)\right) ds}d\tau  \notag \\
&&+n_{2}^{0}(x)e^{-\int_{0}^{t}r_{2}^{-}\left( n_{2}(x,s)\right)
ds}\int_{t}^{\infty }\psi _{2}(\tau )\Psi _{2}^{-1}(\tau -t)f_{2}(\tau
-t|x)d\tau ,  \label{den1}
\end{eqnarray}%
\begin{eqnarray}
j_{2}(x,t) &=&\int_{0}^{t}\int_{\mathbb{R}}\psi _{1}(\tau )j_{1}(z,t-\tau
)p(x-z,\tau )dzd\tau  \notag \\
&&+\int_{t}^{\infty }\int_{\mathbb{R}}\psi _{1}(\tau )\Psi _{1}^{-1}(\tau
-t)n_{1}^{0}(z)f_{1}(\tau -t|z)p(x-z,\tau )dzd\tau .  \label{den2}
\end{eqnarray}%
Balance equations for $n_{1}(x,t)$ and $n_{2}(x,t)$ can be found by
substitution of (\ref{sol1})-(\ref{sol4}) and (\ref{in}) into (\ref{den0})
\begin{eqnarray}
n_{1}(x,t) &=&\int_{0}^{t}\int_{\mathbb{R}}\Psi _{1}(\tau )j_{1}(z,t-\tau
)p(x-z,\tau ))dzd\tau  \notag \\
&&.+\int_{t}^{\infty }\int_{\mathbb{R}}\Psi _{1}(\tau )\Psi _{1}^{-1}(\tau
-t)n_{1}^{0}(z)f_{1}(\tau -t|z)p(x-z,\tau )dzd\tau ,  \label{den3}
\end{eqnarray}%
\begin{eqnarray}
n_{2}(x,t) &=&\int_{0}^{t}\Psi _{2}(\tau )j_{2}(x,t-\tau )e^{-\int_{t-\tau
}^{t}r_{2}^{-}\left( n_{2}(x,s)\right) ds}d\tau  \notag \\
&&+n_{2}^{0}(x)e^{-\int_{0}^{t}r_{2}^{-}\left( n_{2}(x,s)\right)
ds}\int_{t}^{\infty }\Psi _{2}(\tau )\Psi _{2}^{-1}(\tau -t)f_{2}(\tau
-t|x)d\tau .  \label{den4}
\end{eqnarray}%
Note that similar equations have been formulated (not derived) in \cite{FM1}%
. For nonhomogeneous case such as (\ref{L1}) instead of $p(x-z,\tau )$ we
should use $p(x,\tau |z)$. In what follows we use the $\delta $-function for
residence time distribution at $t=0:$
\begin{equation}
f_{i}(\tau |x)=\delta (\tau )\qquad i=1,2.  \label{in2}
\end{equation}%
It corresponds to the case when the residence time of all particles at $t=0$
equals to zero. Substitution of (\ref{in2}) into (\ref{den1})-(\ref{den4})
and rearrangement of the integration variables, $t-t^{\prime }=\tau ,$ give
\begin{eqnarray}
j_{1}(x,t) &=&\int_{0}^{t}\psi _{2}(t-t^{\prime })j_{2}(x,t^{\prime
})e^{-\int_{t^{\prime }}^{t}r_{2}^{-}\left( n_{2}(x,s)\right) ds}dt^{\prime }
\notag \\
&&+n_{2}^{0}(x)\psi _{2}(t)e^{-\int_{0}^{t}r_{2}^{-}\left( n_{2}(x,s)\right)
ds},  \label{den10}
\end{eqnarray}%
\begin{eqnarray}
j_{2}(x,t) &=&\int_{0}^{t}\int_{\mathbb{R}}\psi _{1}(t-t^{\prime
})j_{1}(z,t^{\prime })p(x-z,t-t^{\prime })dzdt^{\prime }  \notag \\
&&+\psi _{1}(t)\int_{\mathbb{R}}n_{1}^{0}(z)p(x-z,t)dz.  \label{den11}
\end{eqnarray}%
The balance equations for the densities $n_{i}(x,t)$ are
\begin{eqnarray}
n_{1}(x,t) &=&\int_{0}^{t}\int_{\mathbb{R}}\Psi _{1}(t-t^{\prime
})j_{1}(z,t^{\prime })p(x-z,t-t^{\prime }))dzdt^{\prime }  \notag \\
&&+\Psi _{1}(t)\int_{\mathbb{R}}n_{1}^{0}(z)p(x-z,t)dz,  \label{den12}
\end{eqnarray}%
\begin{eqnarray}
n_{2}(x,t) &=&\int_{0}^{t}\Psi _{2}(t-t^{\prime })j_{2}(x,t^{\prime
})e^{-\int_{t^{\prime }}^{t}r_{2}^{-}\left( n_{2}(x,s)\right) ds}dt^{\prime }
\notag \\
&&+n_{2}^{0}(x)\Psi _{2}(t)e^{-\int_{0}^{t}r_{2}^{-}\left( n_{2}(x,s)\right)
ds}.  \label{den13}
\end{eqnarray}%
To obtain the nonlinear Master equations for $n_{1}(x,t)$ and $n_{2}(x,t)$
we differentiate the densities given by (\ref{den12}) and (\ref{den13}) with
respect to time $t$%
\begin{eqnarray*}
\frac{\partial n_{1}}{\partial t} &=&\int_{\mathbb{R}}\Psi
_{1}(0)j_{1}(z,t)p(x-z,0)dz-\int_{0}^{t}\int_{\mathbb{R}}\psi
_{1}(t-t^{\prime })j_{1}(z,t^{\prime })p(x-z,t-t^{\prime }))dzdt^{\prime } \\
&&+\int_{0}^{t}\int_{\mathbb{R}}\Psi _{1}(t-t^{\prime })j_{1}(z,t^{\prime })%
\frac{\partial p(x-z,t-t^{\prime })}{\partial t}dzdt^{\prime }-\psi
_{1}(t)\int_{\mathbb{R}}n_{1}^{0}(z)p(x-z,t)dz \\
&&+\Psi _{1}(t)\int_{\mathbb{R}}n_{1}^{0}(z)\frac{\partial p(x-z,t)}{%
\partial t}dz
\end{eqnarray*}%
and
\begin{eqnarray*}
\frac{\partial n_{2}}{\partial t} &=&j_{2}(x,t)-\int_{0}^{t}\psi
_{2}(t-t^{\prime })j_{2}(x,t^{\prime })e^{-\int_{t^{\prime
}}^{t}r_{2}^{-}\left( n_{2}(x,s)\right) ds}dt^{\prime }-n_{2}^{0}(x)\psi
_{2}(t)e^{-\int_{0}^{t}r_{2}^{-}\left( n_{2}(x,s)\right) ds} \\
&&-r_{2}^{-}\left( n_{2}(x,t)\right) \left[ \int_{0}^{t}\Psi
_{2}(t-t^{\prime })j_{2}(x,t^{\prime })e^{-\int_{t^{\prime
}}^{t}r_{2}^{-}\left( n_{2}(x,s)\right) ds}dt^{\prime }+n_{2}^{0}(x)\Psi
_{2}(t)e^{-\int_{0}^{t}r_{2}^{-}\left( n_{2}(x,s)\right) ds}\right] .
\end{eqnarray*}%
Using (\ref{den10})-(\ref{den13}) together with (\ref{kf}), (\ref{kf0}) and $%
\Psi _{1}(0)=1,$ we can rewrite the last system of equations in a compact
form
\begin{equation}
\frac{\partial n_{1}}{\partial t}=L_{x}n_{1}+j_{1}(x,t)-j_{2}(x,t),
\label{fin1}
\end{equation}%
\begin{equation}
\frac{\partial n_{2}}{\partial t}=j_{2}(x,t)-j_{1}(x,t)-r_{2}^{-}\left(
n_{2}\right) n_{2},  \label{fin2}
\end{equation}%
where the densities $j_{1}(x,t)$ and $j_{2}(x,t)$ describe the flux of
particles between a population of spines and a parent dendrite. Now we need
to express $j_{1}(x,t)$ and $j_{2}(x,t)$ in terms of $n_{2}(x,t)$ and $%
n_{1}(x,t)$ respectively. We obtain (Appendix C)
\begin{equation}
j_{1}(x,t)=\int_{0}^{t}K_{2}(t-t^{\prime })n_{2}(x,t^{\prime
})e^{-\int_{t^{\prime }}^{t}r_{2}^{-}\left( n_{2}(x,s)\right) ds}dt^{\prime
},  \label{e1}
\end{equation}%
\begin{equation}
j_{2}(x,t)=\int_{0}^{t}\int_{\mathbb{R}}K_{1}(t-t^{\prime
})p(x-z,t-t^{\prime })n_{1}(z,t^{\prime })dzdt^{\prime },  \label{e2}
\end{equation}%
where $K_{i}(t)$ is the memory kernel defined by
\begin{equation}
\tilde{K}_{i}(s)=\frac{\tilde{\psi}_{i}(s)}{\tilde{\Psi}_{i}(s)}\qquad i=1,2.
\label{K}
\end{equation}%
The nonlinear Master equations (\ref{fin1}) and (\ref{fin2}) together with
interaction terms (\ref{e1}) and (\ref{e2}) is a generalization of a linear
system of equations obtained in \cite{FM1}. In contrast to the classical
Markovian model (\ref{mean1}) and (\ref{mean2}), the spines-dendrite
interaction terms $j_{1}(x,t)$ and $j_{2}(x,t)$ are not-local in time and
space. The density $j_{1}(x,t)$ describes the average flux of particles from
a population of spines through spine necks into parent dendrite. The
characteristic feature of this flux is that it depends on chemical reactions
through the exponential term $e^{-\int_{t^{\prime }}^{t}r_{2}^{-}\left(
n_{2}(x,s)\right) ds}.$ It means that one can not separate the flux of
particles from spines to a dendrite from the chemical reactions inside the
spines. Similar effects have been found for the reaction-transport systems
in \cite{YH,FI07,A}. This phenomenon does not exist in the Markovian case
for which the memory kernel is delta-function
\begin{equation}
\qquad K_{i}(t)=\gamma _{i}\delta (t)  \label{flux}
\end{equation}%
since $\tilde{K}_{i}(s)=\gamma _{i}.$ In this case it follows from (\ref{e1}%
), (\ref{e2}) and (\ref{flux}) that the fluxes of particles are local:
\begin{equation}
j_{1}(x,t)=\gamma _{2}n_{2}(x,t),\qquad j_{2}(x,t)=\gamma _{1}n_{1}(x,t).
\end{equation}%
Note that the flux of particles from dendrite to spines (\ref{e2}) depends
on the transport process inside dendrite.

\section{ Anomalous advection}

Consider the case when there is no reaction inside the spines: $%
r_{2}^{-}\left( n_{2}\right) =0.$ It has been found \cite{FM1} that if the
particle inside dendrite performs a Brownian motion and the residence time's
PDF $\psi _{2}(\tau )$ behaves like
\begin{equation}
\qquad \psi _{2}(\tau )\sim \left( \frac{\tau _{2}}{\tau }\right) ^{1+\mu
},\qquad \mu <1  \label{res2}
\end{equation}%
as $\tau \rightarrow \infty ,$ the mean squared displacement, $<x^{2}(t)>,$
exhibits a subdiffusive behavior $t^{\mu }$. The reason for this anomalous
diffusion is that the population of spines acts as a trap of particles \cite%
{san}.

The aim of this section is to show that if the particles move inside a
dendrite with constant velocity $v,$ the mean particle's position $%
\left\langle x(t)\right\rangle $ increases as $t^{\mu }$ (anomalous
advection). For a constant velocity $v,$ the transport operator takes the
form
\begin{equation}
L_{x}n_{1}=-v\frac{\partial n_{1}}{\partial x}.  \label{vel4}
\end{equation}%
First we find the Laplace transform of $\left\langle x(t)\right\rangle $
\begin{equation}
\left\langle x(s)\right\rangle =-i\left( \frac{dn(k,s)}{dk}\right) _{k=0},
\label{lap}
\end{equation}%
where $n(k,s)=n_{1}(k,s)+n_{2}(k,s)$ is the Fourier-Laplace transform of the
total density of particles $n=n_{1}+n_{2}$:

\begin{equation}
n(k,s)=\int_{\mathbb{R}}\int_{0}^{\infty }e^{ikx-st}n(x,t)dtdx.  \label{FL}
\end{equation}%
Applying the Fourier-Laplace transform to (\ref{fin1})-(\ref{e2}) with $%
r_{2}^{-}\left( n_{2}\right) =0$, we obtain
\begin{equation}
n(k,s)=\frac{n_{1}^{0}(k)\left( \tilde{\Psi}_{1}(s-\varphi (k))+\tilde{\psi}%
_{1}(s-\varphi (k))\tilde{\Psi}_{2}(s)\right) +n_{2}^{0}(k)\left( \tilde{\Psi%
}_{2}(s)+\tilde{\psi}_{2}(s)\tilde{\Psi}_{1}(s-\varphi (k))\right) }{1-%
\tilde{\psi}_{2}(s)\tilde{\psi}_{1}(s-\varphi (k))}.  \label{total}
\end{equation}%
Consider the particular case when the initial density of particles inside
spines is zero ($n_{2}^{0}(k)=0$) and the initial density inside a dendrite
is delta-function: $n_{1}(x,0)=\delta (x)$\ ($n_{1}^{0}(k)=1$). Then
\begin{equation}
n(k,s)=\frac{\tilde{\Psi}_{1}(s-ikv)+\tilde{\psi}_{1}(s-ikv)\tilde{\Psi}%
_{2}(s)}{1-\tilde{\psi}_{2}(s)\tilde{\psi}_{1}(s-ikv)}.  \label{density}
\end{equation}%
since $\varphi (k)=ikv$ for the transport operator defined by (\ref{vel4}).
We assume that the residence time's PDFs $\psi _{1}(\tau )$ is exponential:
\begin{equation*}
\qquad \psi _{1}(\tau )=\gamma _{1}e^{-\gamma _{1}\tau }
\end{equation*}%
with the Laplace transform%
\begin{equation}
\qquad \tilde{\psi}_{1}(s)=\frac{\gamma _{1}}{\gamma _{1}+s}  \label{res5}
\end{equation}%
and $\tilde{\Psi}_{1}(s)=1/(\gamma _{1}+s).$ The Laplace transform $\tilde{%
\psi}_{2}(s)$ corresponding to (\ref{res2}) can be approximated by
\begin{equation}
\qquad \tilde{\psi}_{2}(s)\sim 1-\left( \tau _{2}s\right) ^{\mu },\qquad \mu
<1  \label{res3}
\end{equation}%
for small $s$ \cite{mk}; $\tau _{2}$ is a parameter with units of time. The
mean waiting time $<\tau >=\int_{0}^{\infty }\tau \psi _{2}(\tau )d\tau $ is
infinite in this case. In the limit $s\rightarrow 0,$ we find from (\ref%
{density}), (\ref{res5}) and (\ref{res3}) that%
\begin{equation}
n(k,s)\sim \frac{\gamma _{1}\tau _{2}^{\mu }}{s^{1-\mu }(-ikv+\gamma
_{1}\tau _{2}^{\mu }s^{\mu })}.
\end{equation}%
By using (\ref{lap}) we find

\begin{equation*}
\left\langle x(s)\right\rangle \sim \frac{v}{s^{1+\mu }\gamma _{1}\tau
_{2}^{\mu }}.
\end{equation*}%
Thus the average position of particle is%
\begin{equation*}
\left\langle x(t)\right\rangle \sim \frac{v}{\Gamma (1+\mu )\gamma _{1}\tau
_{2}^{\mu }}t^{\mu },\qquad \mu <1
\end{equation*}%
which is sublinear. This anomalous advection reflects the memory effect
corresponding to the slow movement of particles in spiny dendrites due to
the power law of the residence time distribution for spines (\ref{res2}).

\section{Evolution equation for total density in the long-time limit}

In this section we derive a governing equation for the total density of
particles $n(x,t)=n_{1}(x,t)+n_{2}(x,t)$ in the limit $t\rightarrow \infty $%
. We consider the case when initially all particles are inside dendrite: $%
n(x,0)=n_{1}^{0}(x)$ and $n_{2}^{0}(x)=0$. By using $\tilde{\psi}_{1}(s+ikv)/%
\tilde{\Psi}_{1}(s+ikv)=\gamma _{1},$ we rearrange (\ref{total}) as
\begin{equation}
sn(k,s)-n_{1}^{0}(x)=\frac{\varphi (k)n(k,s)}{1+\gamma _{1}\tilde{\Psi}%
_{2}(s)}.  \label{dens1}
\end{equation}%
Inverse Fourier-Laplace transform gives
\begin{equation}
\frac{\partial n}{\partial t}=\int_{0}^{t}G(t-s)L_{x}n(x,s)ds,
\end{equation}%
where the memory kernel $G$ defined by its Laplace transform
\begin{equation}
\tilde{G}(s)=-\frac{1}{1+\gamma _{1}\tilde{\Psi}_{2}(s)}.
\end{equation}%
Note that this kernel is different from the standard one in CTRW models (\ref%
{K}).

\textit{Standard advection-diffusion equation.} First let us consider the
case when the residence time PDF for spines is
\begin{equation}
\qquad \psi _{2}(\tau )=\gamma _{2}e^{-\gamma _{2}\tau }  \label{res6}
\end{equation}%
with $\tilde{\Psi}_{2}(s)=1/(\gamma _{2}+s)$. We assume that
\begin{equation}
\varphi (k)=ivk-Dk^{2}.  \label{cd}
\end{equation}%
In the long-time limit $t\rightarrow \infty $ ($s\rightarrow 0$), we obtain
from (\ref{dens1}) the governing equation for the total density%
\begin{equation}
\frac{\partial n}{\partial t}+v^{\ast }\frac{\partial n}{\partial x}=D^{\ast
}\frac{\partial ^{2}n}{\partial x^{2}},  \label{sta}
\end{equation}%
where
\begin{equation*}
v_{{}}^{\ast }=\frac{\gamma _{2}}{\gamma _{1}+\gamma _{2}}v,\qquad D^{\ast }=%
\frac{\gamma _{2}}{\gamma _{1}+\gamma _{2}}D.
\end{equation*}%
Note that Eq. (\ref{sta}) is valid for any residence times PDF's $\psi
_{1}(\tau )$ and $\psi _{2}(\tau )$ with finite mean residence times $<\tau
_{1}>$ and $<\tau _{2}>$ respectively. The leading order behavior of the
Laplace transforms $\tilde{\psi}_{1}(s)$ and $\tilde{\psi}_{2}(s)$ as $%
s\rightarrow 0$ is
\begin{equation}
\qquad \tilde{\psi}_{i}(s)\sim 1-<\tau _{i}>s,\qquad i=1,2.
\end{equation}%
Then the effective velocity $v_{{}}^{\ast }$and diffusivity $D^{\ast }$ are
\begin{equation}
v_{{}}^{\ast }=\frac{<\tau _{1}>}{<\tau _{1}>+<\tau _{2}>}v,\qquad D^{\ast }=%
\frac{<\tau _{1}>}{<\tau _{1}>+<\tau _{2}>}D.
\end{equation}

\textit{Anomalous advection-diffusion equation.} Now we consider the case
when the residence time PDF $\psi _{2}(\tau )$ behaves like (\ref{res2}).
The infinite mean residence time leads to the anomalous transport of
particles along spiny dendrite. By using (\ref{res3}) we obtain the Laplace
transform of the survival probability%
\begin{equation}
\tilde{\Psi}_{2}(s)=\frac{1-\tilde{\psi}_{2}(s)}{s}=\frac{\tau _{2}^{\mu }}{%
s^{1-\mu }}.  \label{surva}
\end{equation}%
Substitution of (\ref{surva}) into (\ref{dens1}) gives
\begin{equation*}
\gamma _{1}\tau _{2}^{\mu }(s^{\mu }n(k,s)-s^{\mu -1}n_{1}^{0}(x))=\varphi
(k)n(k,s),
\end{equation*}%
where $\varphi (k)$ is defined by (\ref{cd}). We apply the Fourier-Laplace
transform inversion and obtain the fractional advection-diffusion equation%
\begin{equation}
\gamma _{1}\tau _{2}^{\mu }\frac{\partial ^{\mu }n}{\partial t^{\mu }}+v%
\frac{\partial n}{\partial x}=D\frac{\partial ^{2}n}{\partial x^{2}},
\label{frac1}
\end{equation}%
where
\begin{equation*}
\frac{\partial ^{\mu }n}{\partial t^{\mu }}=\frac{1}{\Gamma (1-\mu )}\frac{%
\partial }{\partial t}\int_{0}^{t}\frac{n(x,\tau )}{(t-\tau )^{\mu }}d\tau -%
\frac{n_{1}^{0}(x)}{\Gamma (1-\mu )t^{\mu }}
\end{equation*}%
is the Caputo fractional derivative \cite{MFH}. Note that the fractional
Nernst-Planck equations for ions' movement have been suggested in \cite%
{Henry}. In general, the velocity of particles $v$ is not constant since the
electric force acting on particles (ions) depends on the density of
particles. This dependence is described by the Poisson-Nernst-Planck (PNP)
equations. The challenge is to derive the fractional analog of these
equations from the mesoscopic model. Obviously the fractional equation (\ref%
{frac1}) is just a first step in this direction.

In fact our model could be used for other applications involving two states:
active phase and a quiescent (immobile) phase. The classical Markovian
switching models involve the Poisson processes with exponential distribution
of residence times (see, for example, \cite{H}). The \ non-Markovian
switching model developed here can be useful in biological modeling \cite{H,
H2, M0,M2}, front propagation in systems with aging and random switching in
velocities \cite{M1}, large-scale transport of solutes in fractured rock
involving mobile/immobile transport with power law memory functions \cite%
{Meer}, directed intermittent search on a tree network \cite{Br}.

\section{ Conclusions}

The aim was to give a \textit{mesoscopic }description of the anomalous
transport and reactions of particles in spiny dendrites. We extended
two-state linear model presented in \cite{FM1} and derived the nonlinear
Master equations for the average densities of particles inside spines and a
parent dendrite. As a starting point we used the Markovian model with an
assumption that the transition probabilities depend on the residence time
variable. Motivated by the experiments \cite{san} on anomalous transport of
particles along the dendrite we assumed that the longer a particle survives
inside a spine, the smaller becomes the transition probability from spine to
dendrite. By using power-law residence time distributions for spines we
found that if particles move inside a dendrite with constant velocity $v,$
the mean particle's position $\left\langle x(t)\right\rangle $ increases as $%
t^{\mu }$ with $\mu <1$ (anomalous subadvection). Fractional
advection-diffusion equation\ for the total density of particles was
derived. We showed that the interaction terms describing the flux of
particles between spines and parent dendrite are not-local in time and
space. In particular the average flux of particles from a population of
spines through spine necks into parent dendrite depends on chemical
reactions in spines. It means that one can not separate the flux of
particles from spines to dendrite from the chemical reactions. This
phenomenon does not exist in the Markovian case. The flux of particles from
dendrite to spines is found to depend on the transport process inside
dendrite. These results might have a significant implication for nerve cell
signalling. The main reason for this is that the effective electro-diffusion
of ions inside spiny dendrite can not be separated form reactions inside
spines. An interesting issue is whether this effect has any significant
effect on the transport of $Ca^{2+}$. One can conclude that the exponential
factor $\exp \left( -\int_{t^{\prime }}^{t}r_{2}^{-}\left( n_{2}(x,s)\right)
ds\right) $ in (\ref{e1}) explains the effect of limited diffusion of $%
Ca^{2+}$ along dendrites observed in experiments \cite{san}. The derivation
of anomalous cable theory from the nonlinear Master equations (\ref{fin1})
and (\ref{fin2}) together with interaction terms (\ref{e1}) and (\ref{e2})
will be the subject of future work.

\section*{Acknowledgment}

This research was supported by the grant $\Gamma $K N02-740-11-5172
"Anomalous transport in biological and chemical systems".

\section{Appendix A}

To derive the system of equations (\ref{dif1}) and (\ref{dif2}), we start
with the balance of particles at the point $x$

\textit{For a dendrite}
\begin{equation}
\xi _{1}(x,\tau +h,t+h)=\left( 1-\gamma _{1}(\tau )h-\lambda (x)h\right) \xi
_{1}(x,\tau ,t)+h\int_{\mathbb{R}}\lambda (z)\xi _{2}(z,\tau
,t)w(x-z)dz+o(h).  \label{bal1}
\end{equation}%
\textit{For spines}
\begin{equation}
\xi _{2}(x,\tau +h,t+h)=\left( 1-\gamma _{2}(\tau )h\right) \xi _{2}(x,\tau
,t)-r_{2}^{-}\left( n_{2}\right) \xi _{2}(x,\tau ,t)h+o(h).  \label{bal2}
\end{equation}%
The first equation states that the density of particles $\xi _{1}(x,\tau
+h,t+h)$ at point $x$ with the residence time $\tau +h$ inside dendrite at
time $t+h$ is the sum of the density of particles with the residence time $%
\tau $ at time $t$ multiplied by the survival probability $1-\gamma
_{1}(\tau )h-\lambda (x)h$ and the density of particles that jump from
different positions $z.$ The jump length $x-z$ is distributed according to
the dispersal kernel or jump length PDF $w(x-z)$. The second equation
describes the balance of particles inside spines. The factor $1-\gamma
_{2}(\tau )h$ in the RHS of (\ref{bal2}) is the probability that the
particles make no transition from spines to dendrite during small time
period $(\tau ,\tau +h]$. The last term describes the decrease of the
density $\xi _{2}$ due to the chemical reaction with the rate $%
r_{2}^{-}\left( n_{2}\right) $ that depends of the local density of
particles $n_{2}.$ The advantage to have a Markovian model is that the
balance of particles during time $(t,t+h]$ is independent of what happened
during the previous time interval $(0,t]$.

Subtracting $\xi _{1}(x,\tau ,t)$ from both sides of the balance equation (%
\ref{bal1}) and subtracting $\xi _{2}(x,\tau ,t)$ from (\ref{bal2}),
dividing by $h,$ and letting $h\rightarrow 0$, we obtain the mesoscopic
system of reaction--transport equations (\ref{dif1}) and (\ref{dif2}) with
the transport operator $L_{x}$ given by (\ref{L2}) .

\section{Appendix B}

First we find the solution to (\ref{dif1}). We denote the Fourier transform
of $\xi _{1}(x,\tau ,t)$ by $\hat{\xi}_{1}(k,\tau ,t).$ Applying the Fourier
transform and convolution theorem to (\ref{dif1}), we obtain
\begin{equation}
\frac{\partial \hat{\xi}_{1}}{\partial t}+\frac{\partial \hat{\xi}_{1}}{%
\partial \tau }=\varphi (k)\hat{\xi}_{1}-\gamma _{1}\left( \tau \right) \hat{%
\xi}_{1},  \label{A1}
\end{equation}%
where $\varphi (k)$ is the symbol of the pseudo-differential operator $L_{x}.
$ It means that the Fourier transform of $L_{x}\xi _{1}$ can be written as
\begin{equation}
\mathcal{F}\left[ L_{x}\xi _{1}(x,\tau ,t)\right] =\varphi (k)\hat{\xi}%
_{1}(k,\tau ,t).
\end{equation}%
In particular, for the advection-diffusion operator (\ref{L1}) with the
constant advection velocity $v$, we have
\begin{equation}
\varphi (k)=ivk-Dk^{2}  \label{A2}
\end{equation}%
For the operator (\ref{L2}) with the constant transition rate $\lambda $, we
obtain
\begin{equation}
\varphi (k)=\lambda (\hat{w}(k)-1),
\end{equation}%
where $\varphi (k)$ is the characteristic exponent of compound Poisson
process with intensity $\lambda $ and $\hat{w}(k)$ is the Fourier transform
of jump PDF $w(z)$ \cite{MFH}.

The characteristics of PDE (\ref{A1}) are determined by the equation $d\tau
/dt=1$. Along the straight lines (characteristics) $\tau (t)=t-t_{0}$ and $%
\tau (t)=t+\tau _{0},$ a partial differential equation (PDE) (\ref{A1}) is
reduced to
\begin{equation}
\frac{d\hat{\xi}_{1}(k,\tau (t),t)}{dt}=\varphi (k)\hat{\xi}_{1}-\gamma
_{1}\left( \tau (t)\right) \hat{\xi}_{1}.  \label{A3}
\end{equation}%
This equation has the solution%
\begin{equation*}
\hat{\xi}_{1}(k,\tau (t),t)=\hat{\xi}_{1}(k,0,t_{0})e^{\varphi
(k)(t-t_{0})-\int_{t_{0}}^{t}\gamma _{1}(\tau (s))ds}\qquad for\qquad \tau
(t)=t-t_{0}
\end{equation*}%
and
\begin{equation*}
\hat{\xi}_{1}(k,\tau (t),t)=\hat{\xi}_{1}(k,\tau _{0},0)e^{\varphi
(k)t-\int_{0}^{t}\gamma _{1}(\tau (s))ds}\qquad for\qquad \tau (t)=t+\tau
_{0}.
\end{equation*}%
Using the inverse Fourier transform and the facts that
\begin{equation*}
\mathcal{F}^{-1}\left[ \hat{\xi}_{1}(k,0,t-\tau )e^{\varphi (k)\tau }\right]
=\int_{\mathbb{R}}\xi _{1}(z,0,t-\tau )p(x-z,\tau )dz\qquad for\qquad \tau
<t,
\end{equation*}%
\begin{equation*}
\mathcal{F}^{-1}\left[ \hat{\xi}_{1}(k,\tau -t,0)e^{\varphi (k)t}\right]
=\int_{\mathbb{R}}\xi _{1}(z,\tau -t,0)p(x-z,t)dz\qquad for\qquad \tau >t,
\end{equation*}%
we obtain (\ref{sol1}) and (\ref{sol2}):
\begin{equation*}
\xi _{1}(x,\tau ,t)=e^{-\int_{0}^{\tau }\gamma _{1}(s)ds}\int_{\mathbb{R}%
}\xi _{1}(z,0,t-\tau )p(x-z,\tau )dz\qquad for\qquad \tau <t,
\end{equation*}%
\begin{equation*}
\xi _{1}(x,\tau ,t)=e^{-\int_{\tau -t}^{\tau }\gamma _{1}(s)ds}\int_{\mathbb{%
R}}\xi _{1}(z,\tau -t,0)p(x-z,t)dz\qquad for\qquad \tau >t,
\end{equation*}%
where $p(x,t)$ is the inverse Fourier transform of
\begin{equation}
\hat{p}(k,t)=e^{\varphi (k)t}.  \label{A4}
\end{equation}%
In particular, $p(x,t)$ is the Green function for the Kolmogorov-Feller
equation
\begin{equation*}
\frac{\partial p}{\partial t}=-\lambda p+\lambda \int_{\mathbb{R}%
}p(z,t)w(x-z)dz
\end{equation*}%
with the initial condition $p(x,0)=\delta (x)$ and $\varphi (k)=\lambda (%
\hat{w}(k)-1).$

The equation for the density $\xi _{2}(x,\tau ,t)$
\begin{equation*}
\frac{\partial \xi _{2}}{\partial t}+\frac{\partial \xi _{2}}{\partial \tau }%
=-\gamma _{2}\left( \tau \right) \xi _{2}-r_{2}^{-}\left( n_{2}\right) \xi
_{2}
\end{equation*}%
can be solved by the method of characteristics in the same way. One can
obtain

\begin{equation*}
\xi _{2}(x,\tau (t),t)=\xi _{2}(x,0,t_{0})e^{-\int_{t_{0}}^{t}\gamma
_{2}(\tau (s))ds-\int_{t_{0}}^{t}r_{2}^{-}\left( n_{2}(x,s)\right) ds}\qquad
for\qquad \tau (t)=t-t_{0}
\end{equation*}%
and
\begin{equation*}
\xi _{2}(x,\tau (t),t)=\xi _{2}(x,\tau _{0},0)e^{-\int_{0}^{t}\gamma
_{2}(\tau (s))ds-\int_{0}^{t}r_{2}^{-}\left( n_{2}(x,s)\right) ds}\qquad
for\qquad \tau (t)=t+\tau _{0}.
\end{equation*}%
These solutions can be rewritten in the forms given in (\ref{sol3}) and (\ref%
{sol4}).

\section{Appendix C.}

Multiplying (\ref{den10}) and (\ref{den13}) by $e^{\int_{0}^{t}r_{2}^{-}%
\left( n_{2}(x,s)\right) ds}$ and taking the Laplace transform $\mathcal{L}%
\left\{ f\right\} $, we obtain
\begin{eqnarray*}
\mathcal{L}\left\{ j_{1}(x,t)e^{\int_{0}^{t}r_{2}^{-}\left(
n_{2}(x,s)\right) ds}\right\} &=&\left[ n_{2}^{0}(x)+\mathcal{L}\left\{
j_{2}(x,t)e^{\int_{0}^{t}r_{2}^{-}\left( n_{2}(x,s)\right) ds}\right\} %
\right] \tilde{\psi}_{2}(s), \\
\mathcal{L}\left\{ n_{2}(x,t)e^{\int_{0}^{t}r_{2}^{-}\left(
n_{2}(x,s)\right) ds}\right\} &=&\left[ n_{2}^{0}(x)+\mathcal{L}\left\{
j_{2}(x,t)e^{\int_{0}^{t}r_{2}^{-}\left( n_{2}(x,s)\right) ds}\right\} %
\right] \tilde{\Psi}_{2}(s).
\end{eqnarray*}%
Then
\begin{equation}
\mathcal{L}\left\{ j_{1}(x,t)e^{\int_{0}^{t}r_{2}^{-}\left(
n_{2}(x,s)\right) ds}\right\} =\mathcal{L}\left\{
n_{2}(x,t)e^{\int_{0}^{t}r_{2}^{-}\left( n_{2}(x,s)\right) ds}\right\} \frac{%
\tilde{\psi}_{2}(s)}{\tilde{\Psi}_{2}(s)}.  \label{B2}
\end{equation}%
Inverse Laplace transform gives
\begin{equation}
j_{1}(x,t)e^{\int_{0}^{t}r_{2}^{-}\left( n_{2}(x,s)\right)
ds}=\int_{0}^{t}K_{2}(t-t^{\prime })n_{2}(x,t^{\prime
})e^{\int_{0}^{t^{\prime }}r_{2}^{-}\left( n_{2}(x,s)\right) ds}dt^{\prime },
\label{B3}
\end{equation}%
where $K_{2}(t)$ is the memory kernel defined by
\begin{equation*}
\tilde{K}_{2}(s)=\frac{\tilde{\psi}_{2}(s)}{\tilde{\Psi}_{2}(s)}.
\end{equation*}%
From (\ref{B3}), we get (\ref{e1}).

Now let us find the expression for $j_{2}(x,t)$ in terms of $n_{1}(x,t).$%
Taking the Fourier-Laplace transform of (\ref{den11}) and (\ref{den12}), we
obtain
\begin{equation*}
\tilde{j}_{2}(k,s)=\left( \tilde{j}_{1}(k,s)+\hat{n}_{1}^{0}(k\right) )%
\mathcal{L}\left\{ \psi _{1}(t)\hat{p}(k,t)\right\} ,
\end{equation*}%
\begin{equation*}
\tilde{n}_{1}(k,s)=\left( \tilde{j}_{1}(k,s)+\hat{n}_{1}^{0}(k\right) )%
\mathcal{L}\left\{ \Psi _{1}(t)\hat{p}(k,t)\right\} .
\end{equation*}%
Taking into account (\ref{A4}) and shift theorem, we find
\begin{equation}
\mathcal{L}\left\{ \psi _{1}(t)\hat{p}(k,t)\right\} =\tilde{\psi}%
_{1}(s-\varphi (k)),\qquad \mathcal{L}\left\{ \Psi _{1}(t)\hat{p}%
(k,t)\right\} =\tilde{\Psi}_{1}(s-\varphi (k)).
\end{equation}%
We have%
\begin{equation}
\tilde{j}_{2}(k,s)=\tilde{n}_{1}(k,s)\frac{\tilde{\psi}_{1}(s-\varphi (k))}{%
\tilde{\Psi}_{1}(s-\varphi (k))}=\tilde{n}_{1}(k,s)\tilde{K}_{1}(s-\varphi
(k)).  \label{B6}
\end{equation}%
Inverse Fourier-Laplace transform of (\ref{B6}) gives (\ref{e2}).

\end{document}